\documentclass[12pt]{article}
\usepackage{amsmath,latexsym}
\usepackage{graphicx}
 \def\cen{\centerline}

\begin{document}

\setlength{\unitlength}{1mm}

 \title{Thin Domain Walls in Lyra Geometry}
 \author{\Large $F.Rahaman^*$, $M.Kalam^{**}$ and $R. Mondal^*$}
\date{}
 \maketitle
 \begin{abstract}
                   This paper studies thin domain walls within the
      frame work of Lyra Geometry.  We have considered two models.  First one is the thin domain wall
      with negligible pressures perpendicular and transverse direction to the wall and
      secondly, we take a particular type of thin domain wall where the pressure in the perpendicular
      direction is negligible but transverse pressures are existed.  It is shown that the
      thin domain walls have no particle horizon and the gravitational force due to them is
      attractive.
  \end{abstract}

 \bigskip
 \medskip

 \cen{ \bf 1. INTRODUCTION }

 \bigskip
 \medskip
  \footnotetext{ PACS NOS : 98.80cq; 04.20jb; 04.50 \\
     \mbox{} \hspace{.2in}Key words and phrases  : Thin domain walls, Lyra Geometry, Attractive Gravitational fields.\\
                              $^*$Dept.of Mathematics, Jadavpur University, Kolkata-700 032, India\\
                              $^{**}$Dept.of Physics, N.N.College for Women, Regent Estate, Kolkata-700 092, India\\
                              E-Mail:farook\_rahaman@yahoo.com
                              }

    \mbox{} \hspace{.2in} The topological defects namely domain walls,cosmic strings,
    monopoles and textures [Kibble(1976);Vilenkin and Shellard(1994)]
    are formed when the universe underwent a series of phase transitions.
    In particular,the appearance of domain wall is associated with
    the breaking of a discrete symmetry i.e. the vacuum manifold M
    consists of several disconnected components . So, the homotopy
    group $\pi_0(M)$ is non-trivial [$\pi_0(M)\neq 1 $] [Vilenkin and Shellard(1994)].
    Hill,Schram and Fry (1989) has suggested that the formation of
    galaxies are due to domain walls produced during a phase
    transition after the time of recombination of matter and
    radiation .  Recently the study of the domain walls and space
    times associated with them has gained renewed cosmological
    interest due to their application in structure formation in
    the universe [Pando et al(1998);Vilenkin (1983);Sikivie and Ipser (1984);
    Schimidt and Wang (1993);Widraw (1989);Chatterji et al (2000);
    Mukherji (1993). \\
    Since the discovery of general relativity by Einstein there
    have been numerous modification of it. Long ago , since
    1951 , Lyra (1951) proposed an alternating theory of Einstein
    gravity.  He suggested a gauge function into the structureless
    manifold that bears a close resemblance to Weyl's geometry.

 \pagebreak

    In the consecutive investigations, Sen (1957) and  Sen and Dunn (1971)
    proposed a new scalar-tensor theory of gravitation and
    constructed an analog of the Einstein field equation based on
    Lyra geometry, which in normal gauge may be written as

        \begin{equation}
              R_{ik} - \frac{1}{2}g_{ik}R + \frac{3}{2}\phi_i
              \phi_k - \frac{3}{4}g_{ik}\phi_m\phi^m = - 8\pi G T_{ik}
              \label{Eq1}
           \end{equation}

      where $\phi_i$ is the displacement vector and other symbols have their usual meaning
    as in Riemannian geometry.\\
    According to Halford (1970) the present theory predicts the same effects within observational
    limits, as far as the classical solar system tests are concerned, as well as tests based on
    the linearised form of field equations.  Soleng (1987) has pointed out that the constant
    displacement field in Lyra's geometry will either include a creation field and be equal to
    Hoyle's creation field cosmology or contain a special vacuum field which together with the
    gauge vector term may be considered as a cosmological term .  Subsequent investigations were
    done by several authors in scalar tensor theory and cosmology within the framework of Lyra
    geometry [Bharma (1974);Karadi and Borikar(1978);Beesham (1986);Singh and Singh(1991);
    Singh and Desikan(1997);Rahaman et al (2001,2002);Casana et al (2005);
    Rahaman et al (2004)].\\
    In recent, Rahaman (2000,2001,2002,2004) has studied some topological defects within the
    framework of Lyra geometry .\\
    In this work, we shall deal with thin domain wall,assuming the time like displacement
    vector
        \begin{equation}
              \phi_i =[\beta(z,t),0,0,0,0]
              \label{Eq2}
           \end{equation}
    And we are looking forward whether the thin domain wall shows
    any significant properties due the introduction of the gauge
    field in the Riemannian geometry.\\

\bigskip
   \medskip
    \cen{ \bf 2. The models and the Basic equations   }
    \bigskip
    \medskip

 The metric for a plane symmetric space time is taken as
 \begin{equation}
               ds^2=e^A(dt^2-dz^2)-e^C(dx^2+dy^2)
         \label{Eq3}
          \end{equation}
where $A=A(z,t) ; C=C(z,t) $. \\

 \pagebreak

The energy stress components in co-moving coordinates for the
thin domain wall under consideration here are given by
\begin{equation}
T_t^t  = \rho,   T_x^x  = T_y^y = p_1 , T_z^z = 0,  T_t^z = 0 .
         \label{Eq4}
          \end{equation}
where $\rho$ is the energy density of the wall , $p_1$ is the
tension along X and Y directions in the plane of the wall and
pressure in
the perpendicular direction to the wall is negligible . \\
The stress tensors $ T_x^x = T_y^y = p_1 $ corresponding to the
tension of the wall along X and Y directions are assumed to be
zero in the {\bf First case} and $ T_x^x = T_y^y =T_t^t= p_1
=\rho$ in
the {\bf Second case} .\\

The field equations for the metric (2) are
        \begin{equation}
  \frac{e^{-A}}{4}[2A'C'-4C''-3(C')^2]
   +\frac{e^{-A}}{4}[(\dot{C})^2 +
2\dot{A}\dot{C}]-\frac{3}{4}\beta^2 e^{-A} = 8 \pi \rho
         \label{Eq5}
          \end{equation}
\begin{equation}
\frac{e^{-A}}{4}[4\ddot{C} +3\dot{C}^2-2\dot{A}\dot{C}]
+\frac{e^{-A}}{4} [-(C')^2 - 2A'C'] +\frac{3}{4}\beta^2e^{-A}=0
         \label{Eq6}
          \end{equation}
 \begin{equation}
   \frac{e^{-A}}{4}[-2A''-2C''-(C')^2]
   +\frac{e^{-A}}{4}[2\ddot{C}+ 2\ddot{A}+
 (\dot{C})^2]+\frac{3}{4}\beta^2 e^{-A} = 8 \pi p_1
         \label{Eq7}
          \end{equation}
 \begin{equation}
   \frac{1}{2}[\dot{C}'+\dot{C}(A'-C')+\dot{A}C']=0
         \label{Eq8}
          \end{equation}
[ "." and " $ '$ " denotes the differentiation w.r.t. t and z
respectively .]

\bigskip
   \medskip
    \cen{ \bf 3. Solutions   }
    \bigskip
    { \bf Case - I :}   $ p_1 = 0 $

To solve the field equations, we shall assume the separable form
of the metric coefficients as follows :
\begin{equation}
   A = A_1(z) + A_2(t) ;
   C = C_1(z) + C_2(t) ;
         \label{Eq9}
          \end{equation}
From equation (8), by using separable form , we get ,

\begin{equation}
   \frac{C_1' - A_1'}{C_1'} = \frac{\dot{A_2}}{\dot{C_2}}= 1 - m
         \label{Eq10}
          \end{equation}

where $(1-m)$ is the separable constant.

This implies ,
\begin{equation}
   A_1 = m C_1
    \label{Eq11}
          \end{equation}
\begin{equation}
   A_2 = (1 - m)C_2
    \label{Eq12}
          \end{equation}

           \pagebreak

From eqn. (6) and eqn. (7) and by using eqn. (9) and (11), we get

\begin{equation}
 2m{C_1^\prime}^2 - (2m + 2)C_1^{\prime\prime} = 2m \ddot{C_2} + 2m
 \dot{C_2}^2 = n
         \label{Eq13}
          \end{equation}
 ( n being the separation constant ) .

 Solving eqn.(13), we get
\begin{equation}
 C_1 = \frac{1}{b} \ln \sinh (abz)
         \label{Eq14}
          \end{equation}
 where $ a^2 = \frac{n}{2m}$ and $ b = \frac{m}{m+1}$ . \\

 For time part , we get
\begin{equation}
 C_2 = \ln \cosh (at)
         \label{Eq15}
          \end{equation}
So, finally the complete solutions for the metric coefficients
may be expressed in the form
\begin{equation}
  e^A = [\sinh(abz)]^{\frac{m}{b}} [\cosh(at)]^{1-m}
    \label{Eq16}
          \end{equation}
\begin{equation}
   e^C =[\sinh(abz)]^{\frac{1}{b}} [\cosh(at)]
         \label{Eq17}
          \end{equation}
The energy of the wall is
\begin{equation}
  8\pi\rho = \frac{a^2}{[\sinh(abz)]^{\frac{m}{b}}[\cosh(at)]^{1-m}}
   [1+\frac{b}{{\sinh(abz)}^2}-{\coth(abz)}^2]
            \label{Eq18}
          \end{equation}
  Here $\beta^2(z,t)$ takes the following form,
\begin{equation}
  \beta^2(z,t)=\frac{1}{3}[a^2[\coth(abz)]^2-\frac{(2m+2)a^2 b}{{\sinh(abz)}^2}
            - a^2{\tanh(at)}^2-\frac{2(2-m)a^2}{{\cosh(at)}^2}]
            \label{Eq19}
          \end{equation}

\bigskip
    \medskip

 { \bf Case - II :}

 Here we construct another model of a thin domain wall. \\
 We assume that $T_x^x = T_y^y =T_t^t =\rho$ . \\
 In view of the above forms of energy stress tensors and using
 field equations, we find the following solutions \\
\begin{equation}
    e^A = [\sinh(DBz)]^{\frac{m}{B}} [\cosh(Dt)]^{1-m}
       \label{Eq20}
          \end{equation}
\begin{equation}
  e^C =[\sinh(DBz)]^{\frac{1}{B}} [\cosh(Dt)]
         \label{Eq21}
          \end{equation}
where $ D^2=\frac{N}{2(m+2)} ; B=\frac{m+2}{m-1}$ and $ N $=
separation constant . \\
\pagebreak

 The energy of the wall is
\begin{equation}
  8\pi\rho = \frac{D^2}{[\sinh(DBz)]^{\frac{m}{B}}[\cosh(Dt)]^{1-m}}
   [1+\frac{B}{{\sinh(DBz)}^2}-{\coth(DBz)}^2]
            \label{Eq22}
          \end{equation}
  Here $\beta^2(z,t)$ takes the following form,
\begin{equation}
  \beta^2(z,t)=\frac{1}{3}[D^2(2m+1)[\coth(DBz)]^2-D^2(2m+1){\tanh(Dt)}^2
               -\frac{4D^2}{{\cosh(Dt)}^2}]
            \label{Eq23}
          \end{equation}

We see that the above solution take the same form as in {\bf
Case-I} except for $m=1$ . The model in {\bf Case- I }exists for
$ m = 1 $ where as the model in {\bf Case-II} does  exist for $ m \neq 1 $ . \\
Therefore the nature of the solutions of the above two models
should be the same except for $m = 1$.
\bigskip
   \medskip

    \cen{ \bf 4. Discussions :  }
     \medskip

From the results given above, it is evident that at any instant
the domain wall density,$\rho$ decreases with the increase of the
distance from the symmetry plane ( both sides of the symmetry
plane ) and $\rho$ vanishes as $z\rightarrow \pm \infty $ . \\

The general expression for the three  space volume is given by
\begin{equation}
  \sqrt{|g_3|} = [\sinh(abz)]^{\frac{m+2}{2b}} [\cosh(at)]^{\frac{3-m}{2}}
            \label{Eq24}
          \end{equation}
Thus the temporal behaviour would be
\begin{equation}
  \sqrt{|g_3|} \sim [\cosh(at)]^{\frac{3-m}{2}}
            \label{Eq25}
          \end{equation}

If $m > 3$, then the three space collapses . On the other hand
when $m < 3$, there are expansion along Z-direction .

\begin{figure}[htbp]
    \centering
        \includegraphics[scale=.8]{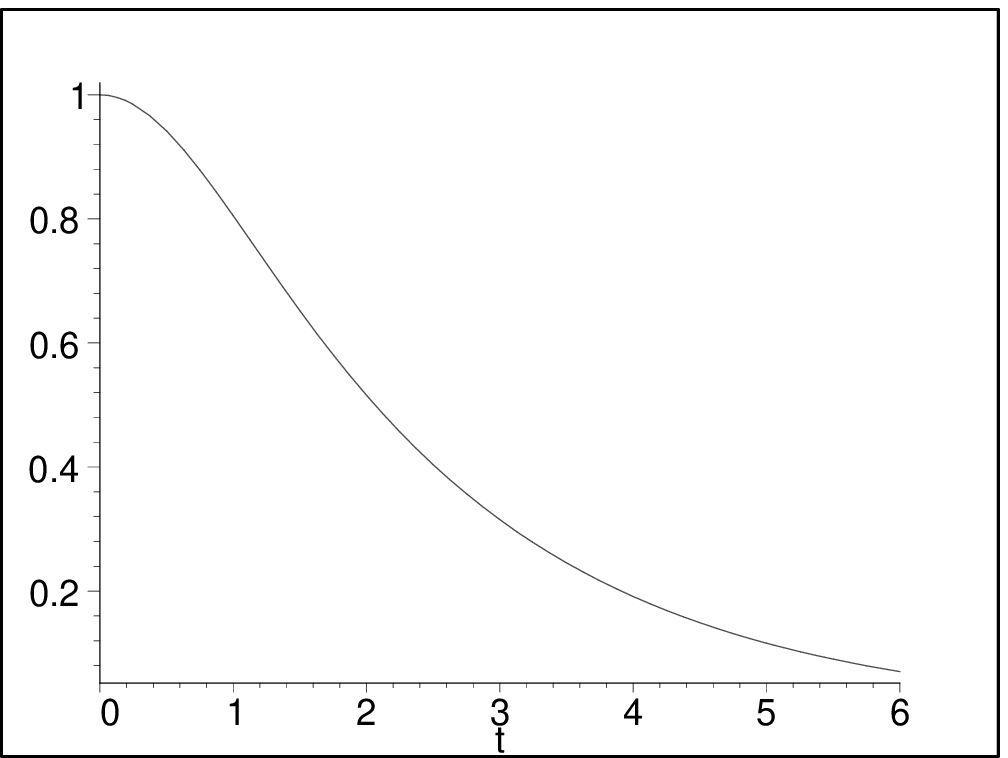}
        \caption{The temporal behaviour of three  space volume for $m>3$}
   \label{fig:fig2}
\end{figure}
\begin{figure}[htbp]
    \centering
        \includegraphics[scale=.8]{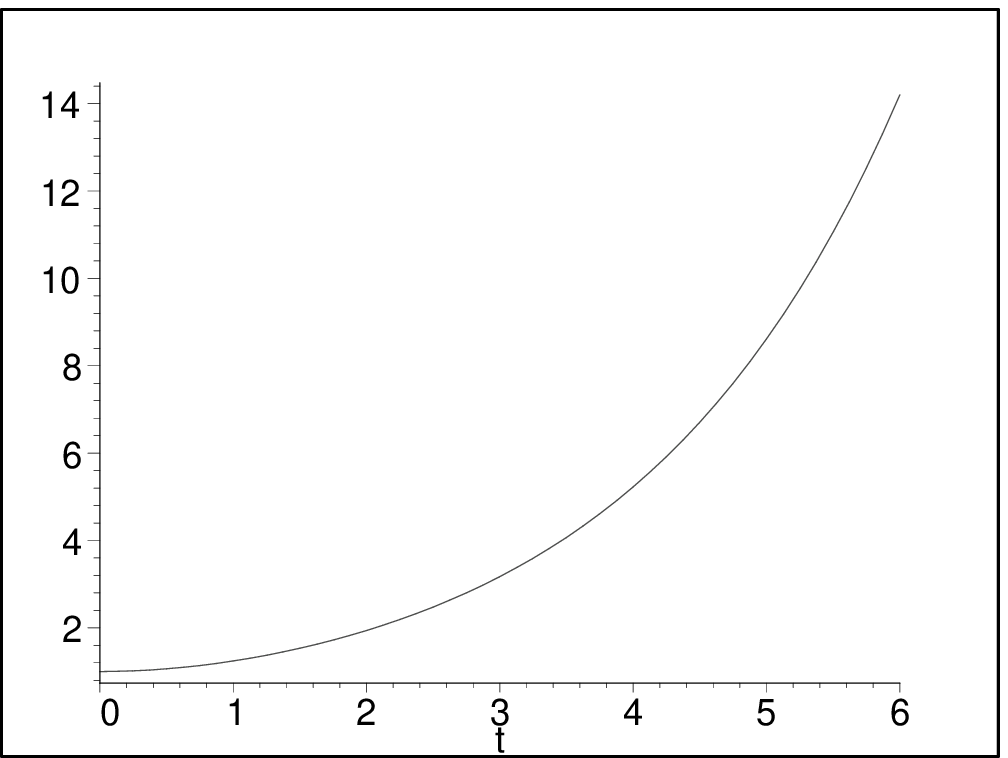}
        \caption{The temporal behaviour of three  space volume for $m<3$}
   \label{fig:fig1}
\end{figure}

\pagebreak

Similar results exist for {\bf Case-II}.

For $ m=1$ , the energy stress components are time independent
whereas metric itself depends on time.
This is similar to Goetz's domain wall [Goetz(1990)].\\
We now calculate the proper distance , $ S_H $( between $z=0$ ,
the centre of the wall and $z\rightarrow\infty$ ) measured along a
space like curve running perpendicular to the
wall($t,x,y=Const.)$ as
\begin{equation}
   S_H =  \int \exp(\frac{A}{2}) dz
       =[\cosh(at)]^{\frac{1-m}{2}}\int
       [\sinh(abz)]^{\frac{m}{2b}} dz
            \label{Eq26}
          \end{equation}
 This distance diverges.\\
 Hence there is no horizon in the Z-direction i.e. perpendicular
 to the wall.This is very similar to the result obtained by
 Wang (Wang, 1994) for a Thick Wall in Einstein's theory but at variance
 with Goetz's result.\\
 The repulsive and attractive character of the wall can be
 discussed by either studying the time like geodesic in the space
 time or analyzing the acceleration of an observer who at rest
 relative to the wall.\\
 Let us consider an observer with four
 velocity given by
 \begin{equation}
   V_i =[\sinh(abz)]^{\frac{m}{2b}}[\cosh(at)]^{\frac{1-m}{2}}\delta_i^t
            \label{Eq27}
          \end{equation}
Then we obtain the Acceleration Vector
\begin{equation}
   A^i=V^i_{;k}V^k=\frac{am}{2}\frac{\coth(abz)}{[\sinh(abz)]^{\frac{m}{b}}
                   [\cosh(at)]^{1-m}}
            \label{Eq28}
          \end{equation}

\bigskip
\medskip
It is evident that $A^i$ is positive and it follows that an
observer who wishes to remain stationary with respect to the wall
must accelerate away from the wall. In other words,the wall
exhibits an attractive nature to the observer . Similar
conclusion can be drawn for the domain wall solution in {\bf Case
- II}. This result is also conformity with Wang [Wang, (1994)] but
differs from Goetz [Goetz(1990)] . We are surprising to note that
the displacement vector still exist after infinite time . For a
future exercise, it will be interesting to study different
properties of different Topological Defects within the frame work
of Lyra Geometry.

\bigskip
 \medskip

        { \bf Acknowledgements }

        We are thankful to the Relativity Cosmology Research Centre, J.U. for helpful discussions.
        We are also grateful to the referee for his constructive suggestions. F.R is thankful to Jadavpur University and DST, Govt. of India for providing financial support
        under Potential Excellence and Young Scientist scheme . \\

\pagebreak
\bigskip
 \medskip

        { \bf References }

      Bharma K. S , Aust. J. Phys. 27, 541 (1974);
                    Karadi T.M and Borikar S.M , Gen Rel. Grav. 1, 431 (1978);
                    Beesham,A  , Ast. Sp. Sc 127, 189 (1986);
                    Singh,T  and  Singh,G.P  , J. Math. Pys. 32, 2456 (1991);
                    Singh,G.P  and Desikan,K  , Pramana  49, 205 (1997);
                    Rahaman F and Bera J, Int.J.Mod.Phys.D10,729 (2001);
                    Rahaman F et al, Astrophys.Space Sci.288,483 (2003);
                    Casana, R ,  Melo,C  and  Pimentel,B  arXiv: gr-qc / 0509096 ;
                    Rahaman F et al, Astrophys.Space Sci.295, 507
                    (2005).

    Goetz,G  , J. Math. Phys. 31, 2683(1990).

    Halford. W ,Aust.J.Phys.23, 833, (1970).

     Hill,C.T. , Schram,D.N. and Fry,J.N.  Nucl. Part. Phys. 19,25 (1989).

       Kibble,T W B  J.Phys.A; Math. and Gen. 9,1387 (1976) .

     Lyra, G   Math. Z 54,52 (1951).

   Pando,J , Valls-Gabant, D and Fang.L  Phys Rev. Lett  81,8568 (1998);
                    Vilenkin, A   Phys. Lett B 133, 177 (1983);
                    Sikivie,P  and Ipser, J   Phys Rev.D 30, 712 (1984);
                    Schimidt,H  and  Wang,A   Phys. Rev. D 47, 4425 (1993);
                    Widraw, L.M  Phys. Rev. D 39, 3571 (1989);
                    Chatterji,S  et al   Grav. and  Cosm. 24, 277 (2000);
                    Mukherji, M   Class. Quan. Grav. 10, 131 (1993).

   Rahaman F, Int.J.Mod.Phys.D9,775 (2000); Rahaman F, Int.J.Mod.Phys.D10,579 (2001);
                    Rahaman F, Astrophys.Space Sci.280,337(2002);
                     Rahaman F et al, Int.J.Mod.Phys.D 10,735 (2001);
                    Rahaman F et al, Fizika B13,719(2004).

   Sen D. K  , Phys. Z 149, 311 (1957);
                    Sen D. K and Dunn K. A , J. Math. Phys 12, 578 (1971);
                   [For brief notes on Lyra geometry see also
                    Beesham, A , Aust. J. Pys. 41, 833 (1988);
                    Singh, T and  Singh,G.P , Int.J.Th.Phys. 31,1433 ( 1992);
                    Matyjasek J, Astrophys.Space Sci.  207,313 (1993)].

     Soleng, H H ,Gen.Rel.Grav. 19,1213 (1987).

   Vilenkin, A and  ShellardE.P.S ,  Cosmic Strings and other Topological
                    Defects (Camb. Univ. Press) (1994).

     Wang . A,  Mod. Phys. Lett. 39, 3605(1994).



\end{document}